\documentclass[superscriptaddress,amsmath,amssymb,amsthm,preprint]{revtex4}
\usepackage{amsmath}
\usepackage{graphicx}
\usepackage{epstopdf}
\usepackage{float}
\usepackage{hyperref}
\usepackage{subfigure}
\usepackage{color}
\usepackage[T1]{fontenc}
\usepackage[utf8]{inputenc}
\usepackage[toc,page]{appendix}
\usepackage[usenames,dvipsnames]{xcolor}
\usepackage[normalem]{ulem}
\usepackage{amssymb}
\usepackage{mathrsfs}
\usepackage{grffile}

\begin{document}

\title{Attractive interactions in the microstructures of asymptotically flat black holes}

\author{Deyou Chen}
\email{deyouchen@hotmail.com}
\affiliation{School of Science, Xihua University, Chengdu 610039, China}

\author{Jun Tao}
\email{taojun@scu.edu.cn}
\affiliation{Center for Theoretical Physics, College of Physics, Sichuan University, Chengdu, 610065, China}

\author{Xuetao Yang}
\email{yangxuetao@stu.scu.edu.cn}
\affiliation{Center for Theoretical Physics, College of Physics, Sichuan University, Chengdu, 610065, China}

\begin{abstract}
In this work, we investigate the microstructure of asymptotically flat black holes with Ruppeiner curvature. Specially, the cosmological constant is considered to have a fluctuation around $0$. Under such consideration, both repulsive and attractive interactions are found in the Reissner-Nordstr\"om and Kerr black holes, while the Schwarzschild black hole has dominant attractive interaction. The result obtained is quite different from that of excluding the fluctuation of cosmological constant, where these black holes are found to be always characterised by repulsive interaction.
\end{abstract}

\maketitle
\newpage

\section{Introduction}\label{section1}
The microstructure of black holes has posed a significant challenge in theoretical physics since the establishment of the four laws of black hole thermodynamics. Alternative avenues of research include exploring macroscopic thermodynamic properties of black holes, which provide valuable insights into the underlying coarse-grained structure. Recently, there has been considerable interest in studying the phase transition of anti de Sitter (AdS) black holes within an extended phase space, where the cosmological constant is treated as a thermodynamic pressure $P$ \cite{Kastor:2009wy, Dolan:2010ha, Dolan:2011xt, Cvetic:2010jb}. This perspective has revealed intriguing similarities between the phase transition of certain AdS black holes and that of a van der Waals (VdW) system \cite{Kubiznak:2012wp, Gunasekaran:2012dq, Kubiznak:2016qmn}, hinting at a fluid-like microstructures for AdS black holes. Various phase transition phenomena observed in AdS black holes further support this notion, including reentrant phase transitions \cite{Altamirano:2013ane}, triple points \cite{Altamirano:2013uqa, Wei:2014hba, Frassino:2014pha}, and superfluid phase transitions \cite{Hennigar:2016xwd}. Other interesting results are found in \cite{GWWY1,GWWY2,GWWY3,GWWY4,GWWY5,RGC,LMLX,CKM,GM,BLT,LW1,LW2,LW3,AMYV,CY,WZ,ACMR1,ACMR2,ACMR3,ACMR4,CHPZ,XCH1,XCH2,XCH3,XCH4,XCH5,XCH6,XCH7,XCH8}.

To further explore microscopic nature of black holes, Ruppeiner geometry has been included. Unlike conventional thermodynamics, this approach reverses the order by utilizing macroscopic thermodynamic details as the foundation for microscopic investigations. In standard thermodynamics, it has been established that the scalar curvature associated with this thermodynamic geometry reflects microscopic interactions, where positive or negative curvature indicates repulsive or attractive interactions, respectively \cite{ruppeiner1979thermodynamics, ruppeiner1981application, janyszek1989riemannian, janyszek1990riemannian, oshima1999riemann, mirza2009nonperturbative, may2013thermodynamic}. Applying this framework to black hole systems is straightforward, as the entropy serves as the thermodynamic potential in Ruppeiner geometry \cite{Ruppeiner:2008kd1,Ruppeiner:2008kd2,Wei:2015iwa,Wei:2019uqg}.

An initial application of this method focused on the Reissner-Nordstr\"om (RN) black hole, revealing a vanishing scalar curvature and suggesting no interaction \cite{Aman:2003ug}. However, considering that the spacetime is curved, it was anticipated that a non-vanishing scalar curvature would be present. This issue was re-evaluated by relating the results of the RN black hole to those of the Kerr-Newman-AdS (KN-AdS) black hole \cite{Mirza:2007ev}. By considering entropy ($S$), angular momentum ($J$), and charge ($Q$) as the complete set of parameter space and appropriately taking the limit for the KN-AdS black hole, a non-vanishing scalar curvature was obtained for the RN black hole. Notably, this scalar curvature is always positive, indicating a dominance of repulsive interaction within the black hole. Additionally, the scalar curvature calculated for the RN black hole in an alternative phase space \cite{Dehyadegari:2016nkd} yielded similar results \cite{Xu:2019nnp}.

In these studies, the cosmological constant is considered as a constant, which can be tuned to 0. Nevertheless, in a more ``fundamental theory'', one may consider the fluctuation of $P$ around $0$, instead of a constant $P=0$ for the asymptotically flat black hole. Equivalently, we can first investigate the microstructure of AdS black holes considering the the fluctuation of $P$, and take the special case of $P\rightarrow 0$ to be the microstructure of asymptotically flat black holes. Indeed, a form of the transition from AdS to asymptotically flat space with a black hole has been proposed in Gauss-Bonnet gravity \cite{Camanho:2013uda,Hennigar:2015mco}. By considering a dynamical variable cosmological constant, i.e. the thermodynamic pressure $P$, it was demonstrated that a phase transition can occur between thermal AdS space and an asymptotically flat geometry with a black hole, regardless of the temperature range. This transition implies that considering the fluctuation of $P$ around $0$ seems to be reasonable in some ``fundamental theory''. Recently, a higher-dimensional origin for the dynamical $P$ was also exactly proposed in the context of holographic braneworlds \cite{Frassino:2022zaz}. It would be interesting to investigate the microstructure of asymptotically flat black hole with the inclusion of the fluctuation of $P$.

In this work, we will regard the cosmological constant to be thermodynamic pressure $P$ that can vary, and investigate the microstructure for RN, Kerr and Schwarzschild black holes by taking  appropriate limits for the Ruppeiner curvature of KN-AdS black hole. As we will see soon, the inclusion of the fluctuation of $P$ in the parameter space will have a significant influence on the microstructure of these black holes.

The paper is structured as follows. Section \ref{section2} provides a concise overview of the thermodynamics of the Kerr-Newmann-AdS black hole in the parameter space of $(S, Q, J, P)$. In Sec. \ref{section3}, the Ruppeiner geometry is constructed for microstructure study. By taking  appropriate limits, the microstructure of RN, Kerr and Schwarzschild black holes are discussed separately. Finally, we present our results in Sec. \ref{section4}.

\section{Thermodynamics for KN-AdS black holes}\label{section2}

For the bulk action of four-dimensional Einstein-Maxwell theory with cosmological constant $\Lambda= -3 / l^{2}$ ($l$ is the AdS radius)
\begin{equation}
\label{eq:1}
S_{E M}=\frac{1}{16 \pi} \int d^{4} x \sqrt{-g} \left(R-F_{\mu \nu} F^{\mu \nu}-2 \Lambda\right),
\end{equation}
the most general black hole solution is the KN-AdS solution that describes a charged, rotating black hole in AdS space. In Boyer-Lindquist-like coordinates, it is given by
\begin{equation}
\label{eq:1}
\begin{aligned}
d s^{2}=&-\frac{\Delta_{r}}{\rho^{2}}\left(\mathrm{~d} t-\frac{a \sin ^{2} \theta}{\Xi} \mathrm{d} \phi\right)^{2}+\frac{\rho^{2}}{\Delta_{r}} \mathrm{~d} r^{2}+\frac{\rho^{2}}{\Delta_{\theta}} \mathrm{d} \theta^{2} \\
&+\frac{\sin ^{2} \theta \Delta_{\theta}}{\rho^{2}}\left(a\mathrm{~d} t-\frac{r^{2}+a^{2}}{\Xi} \mathrm{d} \phi\right)^{2},
\end{aligned}
\end{equation}
where
\begin{equation}
\label{eq:2}
\begin{aligned}
\rho^{2} &=r^{2}+a^{2} \cos ^{2} \theta, \\
\Delta_{r} &=\left(r^{2}+a^{2}\right)\left(1+\frac{r^{2}}{l^{2}}\right)-2 m r+q^{2}, \\
\Delta_{\theta} &=1-\frac{a^{2}}{l^{2}} \cos ^{2} \theta, \\
\Xi &=1-\frac{a^{2}}{l^{2}}.
\end{aligned}
\end{equation}
The associated gauge potential $A$ is given as
\begin{equation}
\label{eq:3}
A=-\frac{q r}{\rho^{2}}\left(\mathrm{~d} t-\frac{a \sin ^{2} \theta}{\Xi} \mathrm{d} \phi\right).
\end{equation}
From these equations, the entropy $S$, temperature $T$, thermodynamic volume $V$, electric potential $\Phi$ and angular velocity of the horizon $\Omega_H$ can be calculated as \cite{Caldarelli:1999xj}
\begin{equation}
\label{eq:4}
\begin{aligned}
S &=\frac{\pi\left(r_{+}^{2}+a^{2}\right)}{\Xi}, \\
T &=\frac{r_{+}\left(1+\frac{a^{2}}{l^{2}}+3 \frac{r_{+}^{2}}{l^{2}}-\frac{a^{2}+q^{2}}{r_{+}^{2}}\right)}{4 \pi\left(r_{+}^{2}+a^{2}\right)}, \\
V &=\frac{2 \pi}{3} \frac{\left(r_{+}^{2}+a^{2}\right)\left(2 r_{+}^{2} l^{2}+a^{2} l^{2}-r_{+}^{2} a^{2}\right)+l^{2} q^{2} a^{2}}{l^{2} \Xi^{2} r_{+}}, \\
\Phi &=\frac{q r_{+}}{r_{+}^{2}+a^{2}}, \\
\Omega_{H} &=\frac{a \Xi}{r_{+}^{2}+a^{2}},
\end{aligned}
\end{equation}
where $r_{+}$ represents the event horizon radius of the black hole. Treating the cosmological constant $\Lambda$ as the thermodynamic pressure $P$ \cite{Kastor:2009wy,Dolan:2010ha,Dolan:2011xt,Cvetic:2010jb},
one can find
\begin{equation}
\label{eq:5}
m=\frac{64 \pi^{2} a^{4} P^{2} q^{2}-48 \pi a^{2} P q^{2}+24 \pi a^{2} P r_{+}^{2}+9 a^{2}+24 \pi P r_{+}^{4}+9 q^{2}+9 r_{+}^{2}}{18 r_{+}}.
\end{equation}
The physical mass $M$, charge $Q$, and angular momentum $J$ can be expressed by
\begin{equation}
\label{eq:6}
M=\frac{m}{\Xi^{2}}, \quad Q=\frac{q}{\Xi}, \quad J=\frac{a m}{\Xi^{2}}.
\end{equation}
After a simple calculation, one can express $T$, $M$ and $V$ in terms of $S$, $J$ and $P$:
\begin{equation}
\label{eq:7}
\begin{aligned}
T &=\frac{-12 \pi^{2} J^{2}+S^{2}\left(64 P^{2} S^{2}+32 P S+3\right)+16 \pi P Q^{2} S^{2}-3 \pi^{2} Q^{4}}{4 \sqrt{\pi} S^{3 / 2} \sqrt{12 \pi^{2} J^{2}(8 P S+3)+\Sigma^{2}}}, \\
M &=\frac{\sqrt{12 \pi^{2} J^{2}(8 P S+3)+\Sigma^{2}}}{6 \sqrt{\pi} \sqrt{S}}, \\
V &=4 \sqrt{S} \frac{2592 \pi^{6} J^{6}(8 P S+3)\left(\pi Q^{2}(8 P S+3)+3 S\right)}{3 \sqrt{\pi} \Sigma\left(36 \pi^{2} J^{2}+\Sigma^{2}\right)^{2} \sqrt{12 \pi^{2} J^{2}(8 P S+3)+\Sigma^{2}}} \\
&+4 \sqrt{S} \frac{432 \pi^{4} J^{4}\left(\pi Q^{2}(8 P S+3)+2 S(4 P S+3)\right) \Sigma^{2}}{3 \sqrt{\pi} \Sigma\left(36 \pi^{2} J^{2}+\Sigma^{2}\right)^{2} \sqrt{12 \pi^{2} J^{2}(8 P S+3)+\Sigma^{2}}} \\
&+4 \sqrt{S} \frac{6 \pi^{2} J^{2} \Sigma^{4}\left(S(8 P S+15)+3 \pi Q^{2}\right)+S \Sigma^{6}}{3 \sqrt{\pi} \Sigma\left(36 \pi^{2} J^{2}+\Sigma^{2}\right)^{2} \sqrt{12 \pi^{2} J^{2}(8 P S+3)+\Sigma^{2}}},
\end{aligned}
\end{equation}
where
\begin{equation}
\label{eq:8}
\Sigma=S(8 P S+3)+3 \pi Q^{2}.
\end{equation}
By identifying the black hole mass as the
enthalpy, the first law of thermodynamics reads
\begin{equation}
\label{eq:9}
d M=T d S+ \Phi d Q +\Omega d J+V d P.
\end{equation}
It will reduce to the RN-AdS and Kerr-AdS black hole cases with $J\rightarrow0$ and $Q\rightarrow0$, respectively. Sequentially, by taking $P\rightarrow0$, the thermodynamics discussed above will reduce to the RN and Kerr black hole cases. If we simultaneously set $J$, $Q$ and $P\rightarrow0$, the KN-AdS black hole will finally degrade into a Schwarzschild black hole.

\section{Ruppeiner geometry}\label{section3}

In this section we explore the microstructure of the RN and Kerr black holes using Ruppeiner geometry. In this thermodynamic geometry, the entropy $S$ is treated as the thermodynamic potential and its fluctuation $\Delta S$ is connected to the line element in the Riemannian geometry \cite{Ruppeiner:1995zz}
\begin{equation}
\label{eq:10}
\Delta l^{2}=-\frac{\partial^{2} S}{\partial x^{\mu} \partial x^{\nu}} \Delta x^{\mu} \Delta x^{\nu},
\end{equation}
where $x^{\mu}$ are the independent thermodynamic quantities.

Within this framework, an empirical observation suggests that the scalar curvature $R$ calculated using Eq. (\ref{eq:10}) is associated with the microscopic interactions occurring in the thermodynamic system under consideration \cite{ruppeiner1979thermodynamics,ruppeiner1981application,janyszek1989riemannian,janyszek1990riemannian,oshima1999riemann,mirza2009nonperturbative,may2013thermodynamic}. Specifically, the positive (negative) sign of $R$ indicates a repulsive (attractive) interaction, while $R=0$ corresponds to the absence of interaction.

Considering the first law of the black hole Eq. (\ref{eq:9}), the basic thermodynamic differential relation between entropy and other thermodynamic quantities can be cast into
\begin{equation}
d S=\frac{1}{T}d M - \sum_{i} \frac{y_{i}}{T} d x^{i},
\end{equation}
where
\begin{equation}
y_{i}=(\Phi,\Omega,V), \quad x^{i}=(Q,J,P).
\end{equation}
With these symbols, the line element Eq. (\ref{eq:10}) can be expressed as
\begin{equation}
\Delta l^{2}=-\Delta z_{\mu} \Delta x^{\mu},
\end{equation}
where $z_{\mu}=(1/T,-y_{i}/T)$ and $x^{\mu}=(M,x^{i})$, and we have the following relationships
\begin{equation}
\begin{gathered}
\Delta z_{0}=\Delta\left(\frac{1}{T}\right)=-\frac{1}{T^{2}} \Delta T, \\
\Delta z_{i}=\Delta\left(-\frac{y_{i}}{T}\right)=\frac{y_{i}}{T^{2}} \Delta T-\frac{1}{T} \Delta y_{i}.
\end{gathered}
\end{equation}
The line element can be then obtained as
\begin{equation}
\Delta l^{2}=\frac{1}{T} \Delta T \Delta S+\frac{1}{T} \Delta y_{i} \Delta x^{i}.
\end{equation}
In this paper, we focus on the fluctuation of $S$ and $P$, thereby the above line element finally becomes
\begin{equation}
\label{eq:11}
\Delta l^{2}=\frac{1}{C_{P}} \Delta S^{2}+\frac{2}{T}\left(\frac{\partial T}{\partial P}\right)_{S} \Delta S \Delta P+\frac{1}{T}\left(\frac{\partial V}{\partial P}\right)_{S} \Delta P^{2}.
\end{equation}
where we have expanded $\Delta T$ and $\Delta P$ by use of $S$ and $P$, and utilized the Maxwell equation $(\partial T / \partial P)_{S}=(\partial V / \partial S)_{P}$ deduced by the first law of thermodynamics Eq. (\ref{eq:9}). Here $C_{P}$ is the heat capacity at constant pressure
\begin{equation}
\label{eq:12}
C_{P}=T\left(\frac{\partial S}{\partial T}\right)_{P}.
\end{equation}

The line element (\ref{eq:11}) was first presented in \cite{Xu:2020gud,Ghosh:2019pwy}
and used to study the thermodynamic geometry of Schwarzschild and charged Gauss-Bonnet black holes in AdS space. In the following, we adopt the approach from \cite{Dolan:2011xt,Wei:2019uqg} and order $P\to 0$ in Eq. (\ref{eq:11}) to investigate the microstructure of RN, Kerr and  Schwarzschild black holes. It is worth noting that the line element (\ref{eq:11}) is not guaranteed to be positive definite, thus the standard thermodynamical stability criterions Eqs. (22)-(24) in \cite{Ruppeiner:2013yca} may not be satisfied, and the thermodynamical stability will be uniquely examined by using the sign of $C_{P}$ in this paper.

\subsection{Microstructure of RN black holes}

\begin{figure}[t]
\centering
\begin{minipage}[b]{1\linewidth}
\centering
\includegraphics[height=9cm]{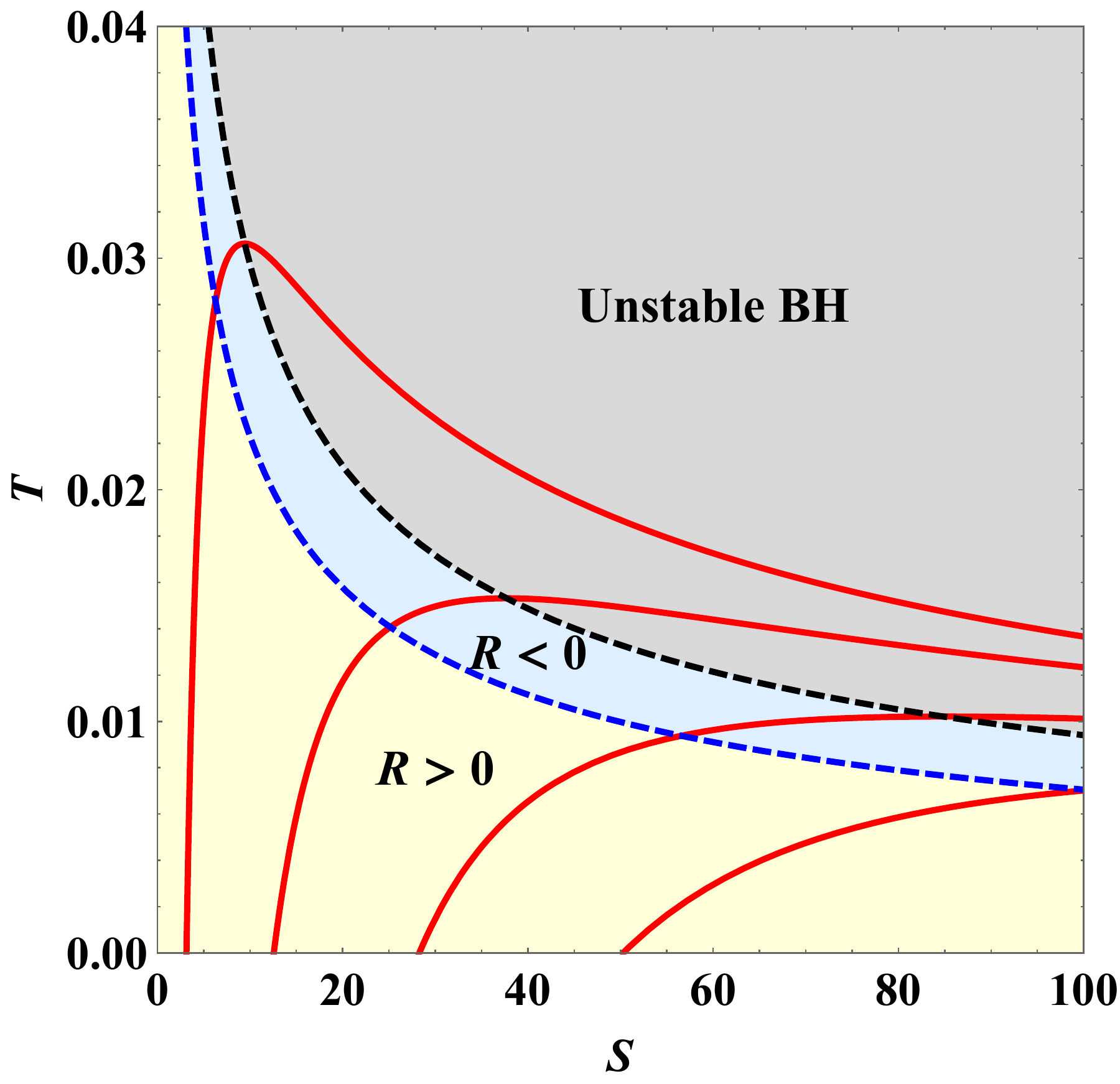}
\label{fig:Fig.1}
\end{minipage}
\caption{\label{fig:Fig.1} The isobaric curves ($P=0$) (red solid curves), spinodal curve (blue dashed line), and sign-changing curve (black dashed line) for the RN black hole in the $T-S$ diagram. The charge for isobaric curves are set as $Q=1$, $2$, $3$ and $4$ from left to right.}
\end{figure}

If we consider the RN systems as the limit of KN-AdS black holes by setting $P\rightarrow0$ and $J\rightarrow0$, the associated thermodynamic curvature will be given by
\begin{equation}
\label{eq:13}
R_{RN}=-\frac{S-2 \pi Q^{2}}{S\left(S-\pi Q^{2}\right)}.
\end{equation}
It diverges at $S=\pi Q^{2}$. Setting the numerator equals zero, we get one zero point
\begin{equation}
\label{eq:14}
S=2 \pi Q^{2}.
\end{equation}
Obviously, $R_{RN}$ takes positive values when $\pi Q^{2}<S<2 \pi Q^{2}$, otherwise takes negative values for $S<\pi Q^{2}$ or $S>2 \pi Q^{2}$. To assess the thermodynamic permissibility of these regions, we plot the isobaric curves ($P=0$) (red solid lines), spinodal curve (black dashed line) and sign-changing curve (blue dashed line) in the $T-S$ diagram, as shown in Fig. \ref{fig:Fig.1}. The isobaric curve for a RN black hole is expressed as
\begin{equation}
\label{eq:15}
T_{P=0}=\frac{S-\pi  Q^2}{4 \sqrt{\pi } S^{3/2}}.
\end{equation}
By using of $\left(\partial_{S} T\right)_{P=0}=0$, one can easily get the spinodal curve
\begin{equation}
\label{eq:16}
T_{sp}=\frac{1}{6 \sqrt{\pi S}}.
\end{equation}
As shown in Fig. \ref{fig:Fig.1}, the spinodal curve separate the black hole phases into stable ($C_{P=0}>0$) and unstable ($C_{P=0}<0$) ones. The unstable phases corresponds to a negative  heat capacity at constant pressure $C_{P=0}=T\left(\partial_{T} S\right)_{P=0}$, which is above the spinodal curve.

From Eqs. (\ref{eq:13}) and (\ref{eq:15}), we can obtain two sign-changing curves in the $T-S$ diagram
\begin{equation}
\label{eq:17}
T_{1}=0, \quad T_{2}=\frac{1}{8 \sqrt{\pi S}}.
\end{equation}
Considering the non-negative requirement of temperature, there is only one physical sign-changing curve $T_{sc}\equiv T_{2}$. The positive $R_{RN}$ region has been labeled in lightyellow color, which is below the sign-changing curve $T_{sc}$, indicating a possible dominant of repulsive interaction among microscopic ingredients of the black hole. When $T_{sc}<T<T_{sp}$, there is a stable phase for negative $R_{RN}$, which may correspond to a attractive interaction. For $T>T_{sp}$, there is another region with negative $R_{RN}$, but such region is thermodynamic unstable.

This result is interesting. Early study \cite{Aman:2003ug} suggested that the RN black hole has a zero thermodynamic curvature. While in Ref. \cite{Mirza:2007ev}, $(S, J, Q)$ was regarded as the complete set of parameter space. By setting cosmological constant $\Lambda \rightarrow0$ and $J\rightarrow0$ in the Ruppeiner curvature calculated for KN-AdS case, the sign-changing behaviour was found only at the extremal limits ($T=0$), where a black hole changes its nature to a new phase, i.e., naked singularity. When $T>0$, the corresponding scalar curvature is always positive, indicating a repulsive-dominanted RN black hole. A similar result was also obtained in Ref. \cite{Xu:2019nnp}, where $(S, q^{2})$ was taken as complete parameter space of a RN black hole.

The difference in these results seems arise from the choice of parameter space. In our case, the $(S, J, Q, P)$ are together taken as complete set of parameter space. Especially, the pressure $P$, i.e., the cosmology constant $\Lambda$ is treated as a parameter that can vary. Then the Ruppeiner curvature for a RN black hole is calculated by taking the limits $P\rightarrow0$ and $J\rightarrow0$ in the Ruppeiner curvature calculated for the KN-AdS case. While in Refs. \cite{Mirza:2007ev,Xu:2019nnp}, the pressure $P$ is considered as a constant, although it can also be tuned to an infinitesimal value.

\subsection{Microstructure of Kerr black holes}
Using the the line element for the complete phase space of parameters ($S,J,Q,P$), and taking the limits $P\rightarrow 0$ and $Q\rightarrow 0$, the Ruppeiner curvature for Kerr black hole can be calculated as
\begin{equation}
\label{eq:18}
R_{Kerr}=\frac{A}{S \left(S^2-4 \pi ^2 J^2\right) B^2},
\end{equation}
where
\begin{equation}
\label{eq:19}
\begin{aligned}
A &=110592 \pi ^{12} J^{12} S^2+118272 \pi ^{10} J^{10} S^4+112000 \pi ^8 J^8 S^6 \\
&\quad +65952 \pi ^6 J^6 S^8+10368 \pi ^4 J^4 S^{10}+8192 \pi ^{14} J^{14}-81 S^{14}, \\
B &=80 \pi ^4 J^4 S^2+72 \pi ^2 J^2 S^4+64 \pi ^6 J^6+9 S^6.
\end{aligned}
\end{equation}
The scalar curvature $R_{Kerr}$ diverges at $S=2 \pi J$. Setting $A=0$, we can get one zero point
\begin{equation}
\label{eq:20}
S\approx 11.7267 J.
\end{equation}
Similar to the discussion in the RN black hole case,  we plot the isobaric curve ($P=0$) (red solid lines), spinodal curves (black dashed line) and sign-changing curve (blue dashed line) in the $T-S$ diagram, as shown in Fig. \ref{fig:Fig.2}. The isobaric curve and spinodal curve for Kerr black holes are calculated as
\begin{equation}
\label{eq:21}
\begin{aligned}
T_{P=0} &=\frac{S^2-4 \pi ^2 J^2}{4 \sqrt{\pi } S^{3/2} \sqrt{4 \pi ^2 J^2+S^2}}, \\
T_{sp} &=\frac{\sqrt{3}+1}{2 \sqrt{2 \left(7 \sqrt{3}+12\right) \pi S} },
\end{aligned}
\end{equation}
which are displayed as red solid line and blue dashed line in Fig. \ref{fig:Fig.2}, respectively. From Eqs. (\ref{eq:18}) and (\ref{eq:21}), we can obtain two sign-changing curves in the $T-S$ diagram
\begin{equation}
\label{eq:22}
T_{1}=0, \quad T_{2}\approx \frac{0.0886341}{\sqrt{S}}.
\end{equation}

\begin{figure}
\centering
\begin{minipage}[b]{1\linewidth}
\centering
\includegraphics[height=9cm]{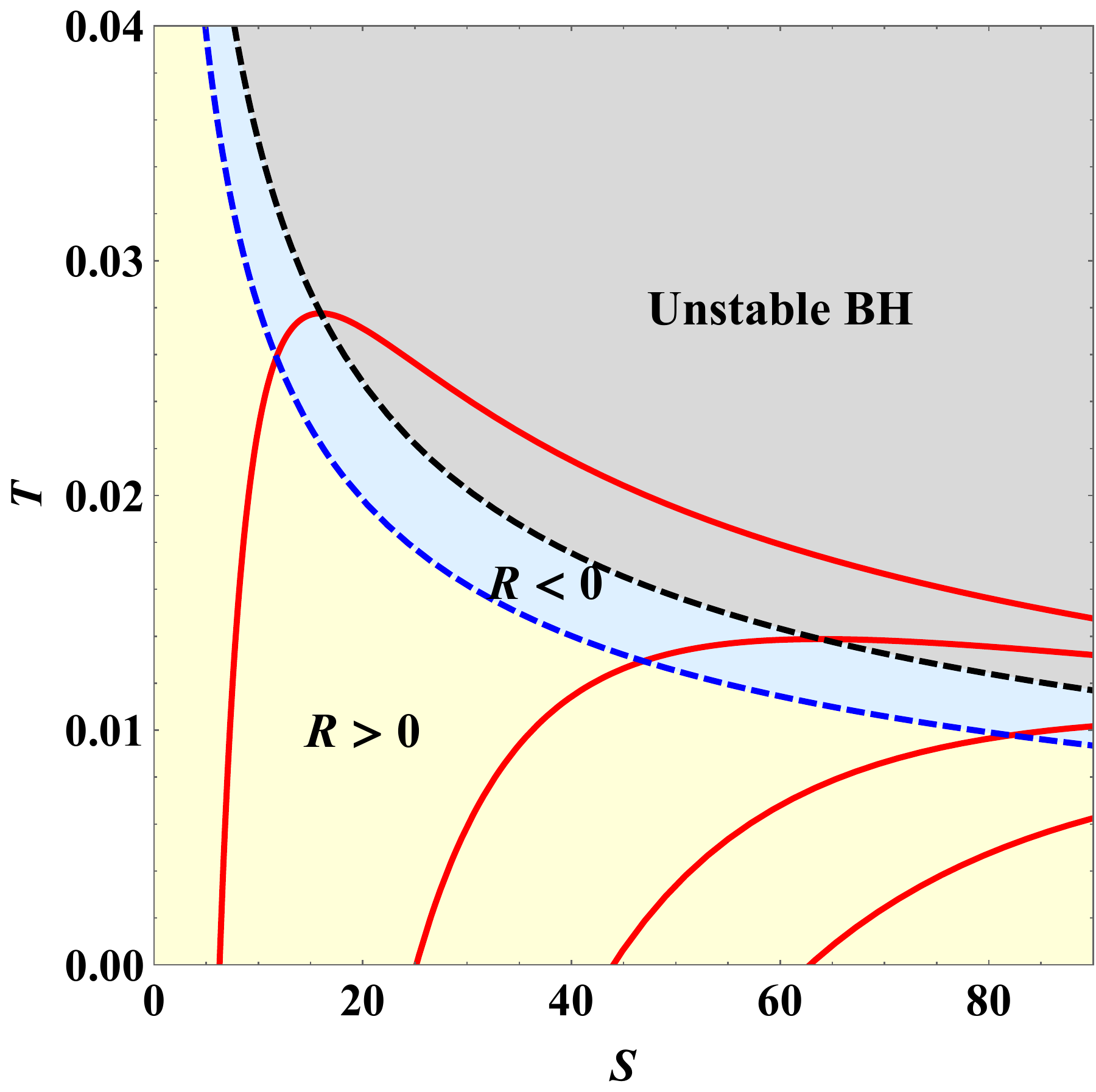}
\label{fig:Fig.2}
\end{minipage}
\caption{\label{fig:Fig.2} The isobaric curves ($P=0$) (red solid curves), spinodal curve (blue dashed line), and sign-changing curve (black dashed line) for Kerr black hole in the $T-S$ diagram. The charge for isobaric curves are set as $Q=1$, $4$, $7$ and $10$ from left to right.}
\end{figure}

Similar to the RN black hole, there is only one physical sign-changing curve $T_{sc}\equiv T_{2}$. The positive $R_{Kerr}$ region is below the sign-changing curve $T_{sc}$, indicating a possible dominant of repulsive interaction among microscopic ingredients of the black hole. When $T_{sc}<T<T_{sp}$, there is a stable phase for negative $R_{Kerr}$, which may correspond to an attractive interaction. In Ref. \cite{Mirza:2007ev}, a scalar curvature for Kerr black hole was also calculated in the $(S, J, Q)$ parameter space. The sign-changing behaviour was found only at the extremal limits ($T=0$), and the corresponding scalar curvature is always positive, implying a repulsive-dominanted Kerr black hole. The difference in results is from the choice of parameter space, as explained before.

\subsection{Microstructure of Schwarzschild black holes}
If we simultaneously set $P$, $J$ and $Q\rightarrow0$ in the Ruppeiner curvature calculated for the KN-AdS case, the curvature for a Schwarzschild black hole will be given as
\begin{equation}
\label{eq:23}
R_{Schwarzschild}=-\frac{1}{S},
\end{equation}
which is always negative. This result suggests an attractive-dominated black hole. The isobaric curve for Schwarzschild black holes is calculated as
\begin{equation}
\label{eq:24}
T_{P=0}=\frac{1}{4 \sqrt{\pi S}},
\end{equation}
which always has negative slope, indicating there is no stable black hole phase with attractive interaction. Interestingly, if the pressure $P$ is not included in the parameter space, one will get an opposite conclusion, where the Ruppeiner curvature is shown to be always positive \cite{Mirza:2007ev,Xu:2019nnp}.

\begin{figure}
\centering
\begin{minipage}[b]{1\linewidth}
\centering
\includegraphics[height=9cm]{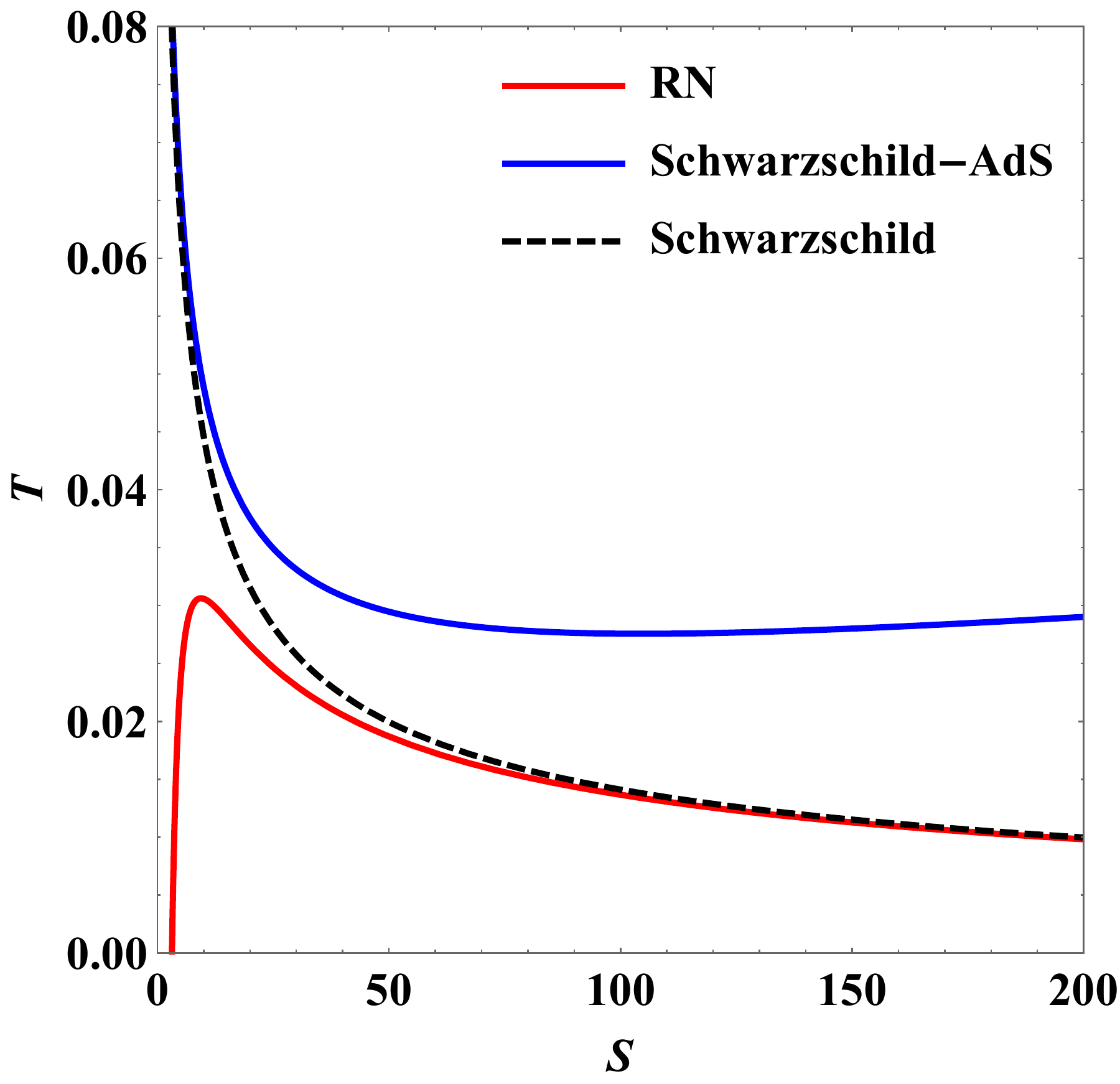}
\label{fig:Fig.3}
\end{minipage}
\caption{\label{fig:Fig.3} The temperature curves for Schwarzschild black hole $(P=Q=J=0)$, RN black hole $(P=J=0,Q=1)$ and Schwarzschild-AdS black hole $(Q=J=0,P=3/(800\pi))$.}
\end{figure}

On the other hand, we notice that there is another interesting conclusion on the microstructure of Schwarzschild black hole: the black hole may be characterised by dominant attraction interaction at low temperature, whereas by dominant repulsive interaction at high temperature \cite{Xu:2020gud}. This conclusion comes from the comparison between temperature curves of Schwarzschild, RN and Schwarzschild-AdS (SAdS) black holes, as we showed in Fig. \ref{fig:Fig.3}. The temperature behavior of Schwarzschild black hole is close to that of SAdS black hole at high temperature (or small scale). The Ruppeiner curvature for SAdS black hole can be calculated by taking the limits $Q\rightarrow 0$ and $J\rightarrow 0$ in KN-AdS case, which turns out to be
\begin{equation}
\label{eq:25}
R_{SAdS}=-\frac{1}{S(1+8 P S)}.
\end{equation}
It always takes negative values. Hence, one can get a conclusion that the Schwarzschild black hole may be characterised by dominant repulsive interaction at high temperature. While at low temperature (or large scale), the temperature behavior of Schwarzschild black hole is close to that of RN black hole. If starting off from a parameter space excluding the pressure $P$, one can say that the Schwarzschild black hole has dominant attraction interaction at low temperature \cite{Mirza:2007ev,Xu:2019nnp}.

Such interaction pattern leaves us an interesting question: which internal mechanism leads to the change in the interaction? However, if the pressure $P$ is included in the parameter space, we can avoid this question, for a simple conclusion that the Schwarzschild black hole is always characterised by dominant attractive interaction at all scale ranges. This conclusion is also self-consistent, since the RN black hole has also dominant attractive interaction at large scale, as discussed before.

\section{Conclusion}\label{section4}
In this paper, we have used the Ruppeiner geometry for KN-AdS black holes to probe the microstructure of RN, Kerr and Schwarzschild black holes, by taking  appropriate limits. We first review the thermodynamics for KN-AdS black holes, where the $(S, J, Q, P)$ are together taken as complete set of parameter space. The Ruppeiner geometry is then constructed to investigate the microscopic interactions. By setting $P\rightarrow0$ and $J\rightarrow0$ in the Ruppeiner curvature calculated for KN-AdS case, the Ruppeiner curvature for a RN black hole is obtained. In the stable black hole phase, both attractive ($R_{RN}<0$) and repulsive ($R_{RN}>0$) interactions are found. Analogously, by setting $P\rightarrow0$ and $Q\rightarrow0$, the Ruppeiner curvature for Kerr black hole is also calculated. Similar to the RN black hole, both attractive and repulsive interactions may exist in the system. If we simultaneously set $P$, $J$ and $Q\rightarrow0$, the Ruppeiner curvature for a Schwarzschild black hole can be easily calculated, which is found to be always negative, indicating an attractive-dominated black hole.

These results are different from the ones obtained from a parameter space excluding the thermodynamic pressure $P$. Specifically, the RN, Kerr and Schwarzschild black holes were found to be repulsive-dominated in the literatures \cite{Mirza:2007ev,Xu:2019nnp}. To conclude, whether regarding the cosmological constant as a dynamical variable, i.e., the pressure $P$, gives quite different microstructures of RN, Kerr and Schwarzschild black holes. In Refs. \cite{Camanho:2013uda,Hennigar:2015mco}, a form of the transition from AdS to asymptotically flat space with a black hole has been proposed. It seems to be reasonable to consider a fluctuation of $P$ around $0$.

\begin{acknowledgments}
We would like to thank Dr. Ningcheng Bai for his useful discussions. This work is supported by NSFC (Grant No. 12275183, 12275184,  12105191).
\end{acknowledgments}

\end{document}